# Ultra-fast treatment plan optimization for volumetric modulated arc therapy (VMAT)

**Chunhua Men,**
*Department of Radiation Oncology, University of California San Diego, La Jolla, CA 92037-0843*

**H. Edwin Romeijn**
*Department of Industrial and Operations Engineering, University of Michigan Ann Arbor, MI 48109-2117*

**Xun Jia, and Steve B. Jiang**[†]
*Department of Radiation Oncology, University of California San Diego, La Jolla, CA 92037-0843*

**Purpose**: To develop a novel aperture-based algorithm for volumetric modulated arc therapy (VMAT) treatment plan optimization with high quality and high efficiency.

**Methods**: The VMAT optimization problem is formulated as a large-scale convex programming problem solved by a column generation approach. We consider a cost function consisting two terms, the first which enforces a desired dose distribution while the second guarantees a smooth dose rate variation between successive gantry angles. At each iteration of the column generation method, a subproblem is first solved to generate one more deliverable MLC aperture which potentially decreases the cost function most effectively. A subsequent master problem is then solved to determine the dose rate at all currently generated apertures by minimizing the cost function. The iteration of such an algorithm yields a set of deliverable apertures, as well as dose rates, at all gantry angles.

**Results**: The algorithm was preliminarily tested on five prostate and five head-and-neck clinical cases, each with one full gantry rotation and without any couch/collimator rotations. Compared to corresponding co-planar IMRT treatment plans (9 fields for prostate cases and 5 fields for head-and-neck cases), the VMAT plans delivered much lower doses to critical structures and more conformal doses to targets. Moreover, extremely high efficiency has been achieved in our algorithm. It takes only 5~8 minutes on CPU (MATLAB code on an Intel Xeon 2.27 GHz CPU) and 18~31 seconds on GPU (CUDA code on an NVIDIA Tesla C1060 GPU card) to generate such a plan.

**Conclusions**: We have developed an aperture-based VMAT optimization algorithm which can generate clinically deliverable and high quality treatment plans at very high efficiency.

Key words: VMAT, plan optimization, column generation method, GPU

---

[†] Electronic mail: sbjiang@ucsd.edu

## 1. Introduction

Volumetric modulated arc therapy (VMAT) is considered as one of the most promising radiotherapy technologies with great potential to improve the treatment quality. Perhaps more importantly, due to its very high treatment delivery efficiency, VMAT has the potential to allow patients to receive treatment in a more timely fashion and also make modern radiotherapy available to more cancer patients in resource-limited regions. Additionally, shorter delivery time indicates reduced probability of treatment errors caused by patient motion during the treatment.

In a VMAT treatment process, treatment gantry rotates around the patient while the radiation beam dynamically changes its aperture shape and associated intensity. By optimizing the beam aperture shape formed by a multi-leaf collimator (MLC) and the beam intensity at each gantry angle, a precisely sculpted desirable 3D dose distribution can be attained. This optimization problem is extremely complicated due to the very large scale of the problem and hardware constraints imposed on neighboring beam apertures and intensities. It can hardly be mathematically modeled in a concise and clean manner. Currently, this problem can only be solved in a brute force way by using simple heuristic algorithms, which usually take up to hours to find a solution and cannot guarantee its optimality[1-16]. The use of such algorithms has considerably limited the exploitation of VMAT's great potentials. In this letter, we present a novel aperture-based algorithm for VMAT treatment plan optimization with high plan quality and computational efficiency.

## 2. Methods and Materials

### 2.1 Optimization model

We denote the number of beams by $N$ and these beams are sorted based on the beam angles from $0^0$ to $359^0$. Note that a beam aperture is a snapshot of the MLC leaf positions at a time point during the radiation dose delivery. Let us decompose each beam aperture into a set of beamlets and denote the set of beamlets exposed in beam $k$ at angle $\theta_k$ by $A_k$. With beam $k$ we associate a decision variable $y_k$ that indicates the intensity of that aperture. The set of voxels that represents the patient's CT image is denoted by $V$. In addition, we denote the dose to a voxel $j$ by $z_j$ ($j \in V$) and it is calculated using a linear function of the intensities of the apertures through the so-called dose deposition coefficients $D_{ij}$, the dose received by the voxel $j \in V$ from the beamlet $i \in A_k$ at unit intensity: $z_j = \sum_{k=1}^{N} y_k \sum_{i \in A_k} D_{ij}$. We calculate $D_{ij}$'s using our in-house dose calculation engine implemented on a general purpose graphics processing unit (GPU)[17].

Our VMAT optimization model employs a cost function with quadratic one-sided voxel-based penalties. Specifically, we write the cost function for a voxel $j \in V$ as:

$$F(z_j) = \alpha_j \big(\max\{0, T_j - z_j\}\big)^2 + \beta_j \big(\max\{0, z_j - T_j\}\big)^2 \tag{1}$$

where $\alpha_j$ and $\beta_j$ represent the weights for underdosing and overdosing penalty, respectively. For target voxels, we set $\alpha_j > 0$ and $\beta_j > 0$ to penalize any deviation from the prescription dose $T_j$. As for critical structures, $\alpha_j = 0$ and $\beta_j > 0$ are chosen to add penalty for only those voxels received dose exceeding a threshold $T_j$.

In a VMAT system, the dose rate variation between neighboring angles is constrained within a certain range. To ensure the plan deliverability regarding this constraint, we add a smoothing term in the cost function to minimize difference between beam intensities at two neighboring beam directions, which is formulated as

$$G(\boldsymbol{y}) = \sum_{k=1}^{N-1}(y_{k+1} - y_k)^2/(\theta_{k+1} - \theta_k). \tag{2}$$

Our VMAT optimization model then can be written as

$$\min_{y, A_k} F(\boldsymbol{z}) + \gamma G(\boldsymbol{y}) \tag{3}$$

subject to     $z_j = \sum_{k=1}^{N} y_k \sum_{i \in A_k} D_{ij}$

$y_k \geq 0 \qquad\qquad\qquad k = 1,2,\dots,N,$

where $\gamma > 0$ is a factor adjusting relative weights between the two terms. Note that $A_k$, the set of beamlets in the aperture of beam $k$, is also a decision variable to be optimized.

### 2.2   Optimization algorithm

A column generation approach is developed to deal with the extremely large dimensionality of the VMAT optimization problem. This method has been successfully used to solve direct aperture optimization (DAO) problem for IMRT treatment planning in our previous studies[18-20]. In a VMAT treatment planning optimization problem, in addition to nonnegative beam intensity constraints and MLC hardware deliverability constraints, as in the DAO problem, there exist constraints due to the complexity of VMAT delivery technique. These are categorized into 1) the maximum leaf motion speed, and 2) the maximum dose rate variation. We have considered the second additional constraint in the objective function as penalty-based soft constraint and here we describe how to handle the first constraint in our column generation method implementation. In our VMAT treatment plan optimization, a single $360^0$ gantry rotation is discretized into uniformly spaced 180 beam directions. Our algorithm optimizes the shape and the intensity of the MLC aperture for each beam direction. The algorithm generates MLC apertures one by one by solving a master problem and a subproblem (also called *pricing* problem) iteratively. The subproblem at each iteration generates the most promising MLC aperture from all un-occupied beam directions, while accounting for all the deliverability constraints imposed by the MLC system, including the aforementioned constraint 1). By checking the KKT optimality conditions[21], we can obtain the "price" for each beamlet[18-20] and we are trying to find a set of beamlets to form an deliverable aperture from all unoccupied beams which has the best total "price". By checking the shapes of existing neighboring MLC apertures, we can find such a most promising aperture at each iteration by passing through all feasible beamlets in a row of a given beam from left to right only once according a polynomial-time algorithm[19]. By limiting to one aperture in each beam direction and taking MLC leaf motion constraints into account, the subproblem has the ability to generate deliverable high-quality apertures. The master problem is to solve our VMAT optimization model (3) with optimized $A_k$. The gradient projection method is used to solve the master problem. The flowchart of column generation method for VMAT treatment plan optimization algorithm is summarized in Figure 1.

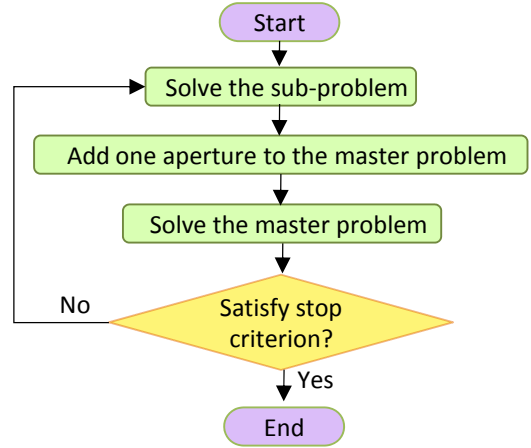

**Figure 1.** A flowchart of our algorithm for solving the VMAT plan optimization problem.

### 2.3　　GPU implementation

GPU offers a potentially powerful computational platform for convenient and affordable high-performance computing and researchers have been starting to use GPU solving heavy duty problems in a clinical context[17, 22-26]. To speed up the VMAT optimization algorithm, we implemented the column generation method on GPU under the Compute Unified Device Architecture (CUDA). The GPU implementation is very similar to our DAO implementation[20], except for more complex MLC constraints are included.

### 3.　　Results and discussion

Five clinical prostate cases (P1-P5) and five clinical head-and-neck cases (H1-H5) were used to evaluate our new algorithm in terms of treatment plan quality and planning efficiency. For prostate cases, the prescription dose to planning target volume (PTV) was 73.8 Gy and for the head-and-neck cases, the prescription dose was 73.8 Gy to PTV1 and 54 Gy to PTV2. PTV1 consists the gross tumor volume (GTV) expanded to account for both sub-clinical disease as well as daily setup errors and internal organ motion; PTV2 is a larger target that also contains high-risk nodal regions and is again expanded for same reasons. For all cases, we used a beamlet size of 10×10 mm$^2$ and voxel size of 2.5×2.5×2.5 mm$^3$ for target and organs at risk (OARs). For unspecified tissue (*i.e.*, tissues outside the target and OARs), we increased the voxel size in each dimension by a factor of 2 to reduce the optimization problem size. The full resolution was used when evaluating the treatment quality (does volume histograms (DVHs), dose color wash, isodose curves, *etc.*). The case dimensions are showed in Table 1.

150　**Table 1.** Case dimensions and CPU/GPU running time on an Intel Xeon 2.27 GHz CPU and an NVIDIA Tesla C1060 GPU for our VMAT plan optimization implementations.

| Case | # beamlets | # voxels | # non-zero $D_{ij}$'s ($\times 10^7$) | CPU time (sec) | GPU time (sec) |
|---|---|---|---|---|---|
| P1 | 40,620 | 45,912 | 2.3 | 340 | 22 |
| P2 | 59,400 | 48,642 | 3.2 | 265 | 18 |
| P3 | 38,880 | 28,931 | 1.8 | 276 | 20 |
| P4 | 43,360 | 39,822 | 2.6 | 410 | 26 |
| P5 | 51,840 | 49,210 | 3.0 | 348 | 23 |
| H1 | 51,709 | 33,252 | 2.5 | 290 | 21 |
| H2 | 78,874 | 59,615 | 5.0 | 468 | 27 |
| H3 | 90,978 | 74,438 | 5.5 | 342 | 25 |
| H4 | 71,280 | 31,563 | 2.6 | 363 | 25 |
| H5 | 53,776 | 42,330 | 3.5 | 512 | 31 |

　　　Figure 2 shows two typical VMAT plans for a prostate case and a head-and-neck case. Compared to corresponding 9-field prostate and 5-field head-and-neck co-planar IMRT plans, our VMAT plans deliver much lower doses to OARs and more uniform
155　doses to the targets. For the head-and-neck case, there are many more critical structures used in the optimization, such as brain stem, optic nerve, spinal cord, *etc*., whose doses are very low and thus DVH curves are not shown in Figure 2 for clarity purpose.

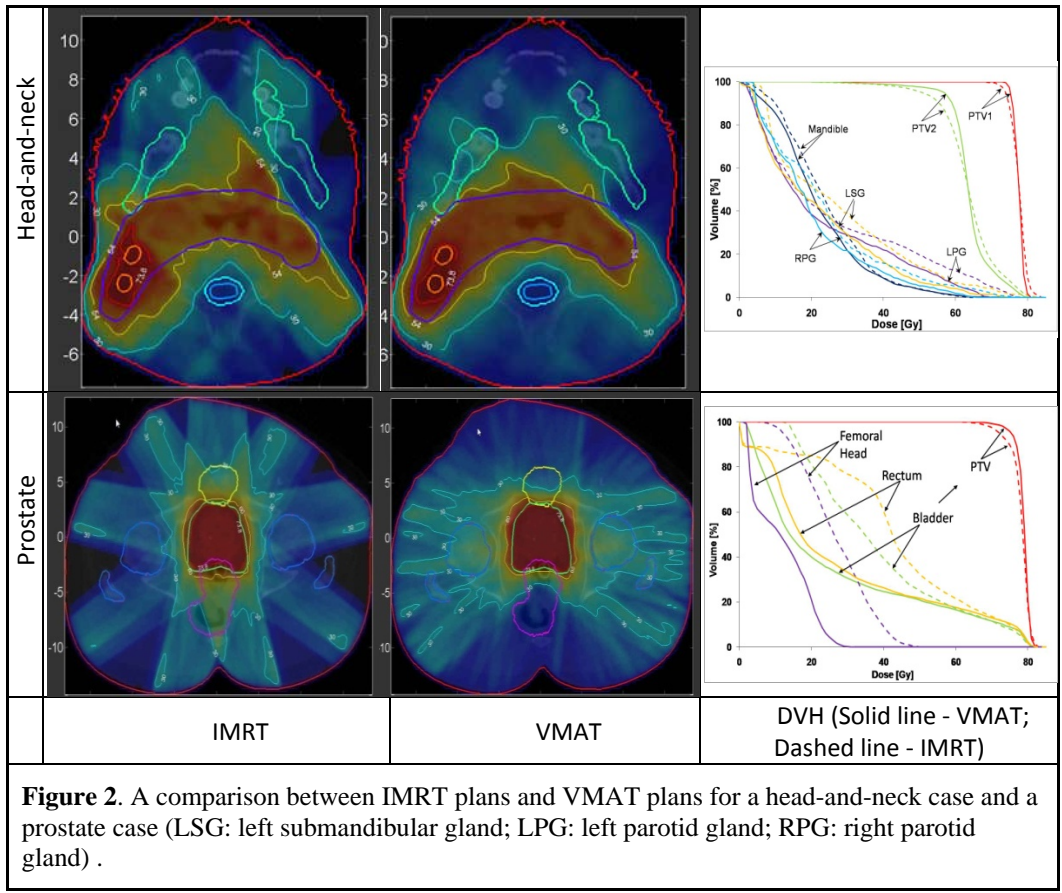

**Figure 2**. A comparison between IMRT plans and VMAT plans for a head-and-neck case and a prostate case (LSG: left submandibular gland; LPG: left parotid gland; RPG: right parotid gland) .

In terms of planning efficiency, to generate such a plan, it only takes 5~8 minutes with the MATLAB implementation on an Intel Xeon 2.27 GHz CPU and 18~31 seconds with the CUDA implementation on an NVIDIA Tesla C1060 GPU card, as shown in Table 1, which makes VMAT a possible treatment delivery technique for online adaptive radiation therapy.

In the optimization model, we add dose rate constraints as penalty-based soft constraints in the cost function. However, these constraints need to be satisfied as hard constraints. Note that we only have to satisfy these dose rate constraints at the last iteration after generating the last aperture. We ensure the plan feasibility only at the last iteration step: if the final solution is not feasible, the coefficient $\gamma$ in the objective function will be automatically increased and then the intensity for each aperture will be re-optimized till the solution is feasible. Similar approach can also be used to handle the maximum and minimum dose rate constraints, which are not considered in this preliminary work.

**Acknowledgements**

We would like to thank NVIDIA for providing GPU cards. This work is supported in part by the University of California Lab Fees Research Program.